# Anisotropic anomalous Hall effect in distorted kagome GdTi$_3$Bi$_4$

*Avdhesh K. Sharma,[1] Bo Tai,[1] Subhajit Roychowdhury,[1, 2] Premakumar Yanda,[1] Ulrich Burkhardt,[1] Xiaolong Feng,[1] Claudia Felser[1, *] and Chandra Shekhar[1, *]*

[1]Max Planck Institute for Chemical Physics of Solids, 01187 Dresden, Germany
[2]Department of Chemistry, Indian Institute of Science Education and Research Bhopal, Bhopal-462 066, India

**Abstract:**

Topological kagome magnets offer a rich landscape for exploring the intricate interplay of quantum interactions among geometry, topology, spin, and correlation. GdTi$_3$Bi$_4$ crystallizes in layered Ti-based kagome nets intertwined with zigzag Gd chains along the *a*-axis and orders antiferromagnetically below ~15 K. Here, we present the temperature and field-dependent electrical transport of GdTi$_3$Bi$_4$ in different directions. The material exhibits anomalous Hall conductivity (AHC) of 410 $\Omega^{-1}$ cm$^{-1}$ at 2 K for $\mu_0 H \parallel c$ and it is completely absent for $\mu_0 H \parallel a$, despite the similar magnetization observed in both orientations. This behavior is quite contradictory, as anomalous Hall effect (AHE) typically scales with the magnetization. Through first-principles calculations, it is demonstrated that in the presence of time-reversal symmetry broken by the Gd-4f sublattice and spin–orbit coupling, the magnetization direction controls the orbital mixing in the Ti $t_{2g}$ bands, relocating Berry-curvature hot spots and producing the observed orientation-selective AHC. The results establish GdTi$_3$Bi$_4$ as platform for investigating new avenues of AHE, such as directional AHE, and thus shed new light on the intricate coupling between magnetic and electronic structures, paving the way for exploring novel quantum phenomena.

………………………….
*Email: felser@cpfs.mpg.de; shekhar@cpfs.mpg.de

**Introduction:**

The intrinsic anomalous Hall effect (AHE) has emerged as a key probe of quantum phenomena in magnetic topological materials, arising from a pseudomagnetic field generated by momentum-driven Berry curvature (BC), which further originates from the topology of the band structure and its invariants.[1-3] A finite net magnetization with time-reversal symmetry (TRS) broken is prerequisite criterion, although AHE in materials with complex magnetic spin textures and non-trivial topological band structures is always promising to investigate.[4-6] A kagome lattice, composed of interconnected triangles arranged in a hexagonal pattern, presents a unique atomic structure where each atom is surrounded by four neighboring atoms within the plane.[7] This distinctive geometry provides a unique platform in which Dirac cones, flat bands, and van Hove singularities naturally coexist, enabling strong source of BC once symmetry is lowered.[8-16]

Over the past few years, several kagome series such as $A$V$_3$Sb$_5$ (where $A$ can be K, Cs, or Rb) and $Ln$M$_6$X$_6$ (where $Ln$ is a rare-earth element, $M$ is a transition metal, and $X$ signifies p-block element) have exhibited conventional and unconventional AHE and other related quantum phenomena due to presence of either finite or inherent magnetic moment.[14,17-23] However, the origin of AHE remains a topic of debate in $A$V$_3$Sb$_5$ series, with loop current proposed as source of AHE in the absence of long range magnetic ordering.[23-25] Among $Ln$M$_6$X$_6$ series, a number of predominantly ferromagnetic (FM) compounds exhibit AHE [26,27] and anomalous Nernst effect in YMn$_6$Sn$_6$ [10], and Chern magnetism in TbMn$_6$Sn$_6$.[15] However, AHE without presence of such magnetism is observed in ScV$_6$Sn$_6$, which persists up to the CDW transition.[11,28] Recently, diverse electronic and magnetic properties have been discovered in titanium-based bilayer kagome metal, $Ln$Ti$_3$Bi$_4$. The crystal structure is formed by slightly distorted Ti-kagome, interleaved with zigzag $Ln$ chains within $Ln$Bi bilayers.[29] It offers a tunable magnetic order through the $Ln$ site, while keeping the kagome-derived electronic states largely intact. Across the series, many physical properties such as spin density waves, van Hove singularities, stripe-like magnetization plateaus, and even charge–spin intertwined density waves have been observed, underscoring the pivotal role of one dimensional magnetic chain of *Ln* and Ti- kagome lattice.[30-38] A key signature of rare-earth elements has been the emergence of magnetization plateaus or step-like transition under the application of magnetic fields. In the case of GdTi$_3$Bi$_4$, this phenomenon occurs as a result of the reconfiguration of localized moments into stripe-like or fractional magnetization states. This supports the formation of unconventional charge-spin-intertwined density wave, which appears to be limited to the *c*-axis. Furthermore, the emergence of two distinct antiferromagnetic (AFM) order states has been observed, exhibiting a deviation of ±7° from the *a*-axis. [32,38] Such field induced textures, sometimes produce substantial BC and AHE, as seen in the rare-earth based compounds.[39-42] However, the manner in which the localized spins of *Ln* influence the electronic bands remains largely unexplored in the contexts of connection between magnetism and topology as reflected in their electrical transport properties.

In this study, we conducted axis dependent magneto-electrical transport investigations on single crystals of GdTi$_3$Bi$_4$ followed by a theoretical analysis. We explicitly observed a large AHC of 410 $\Omega^{-1}$ cm$^{-1}$ at 2 K with $\mu_0 H \parallel c$ and it is completely absent for $\mu_0 H \parallel a$. However, the magnetic behavior is similar in both directions, including 1/3 magnetic plateau and other minor plateaus. Fully relativistic calculations in saturated FM configurations with $\mu_0 H \parallel c$ or $\mu_0 H \parallel a$ reproduce the strong anisotropy and trace it to spin-orbit coupling (SOC)-gapped anticrossings near Fermi energy ($E_F$). Our findings highlight that the intertwining of magnetism and structure, via TRS breaking due to the exchange field of Gd-4f sublattice and orbital mixing controlled by SOC, is vital for elucidating the axis dependent anomalous transport observed in kagome compounds.

**Results and Discussion:**

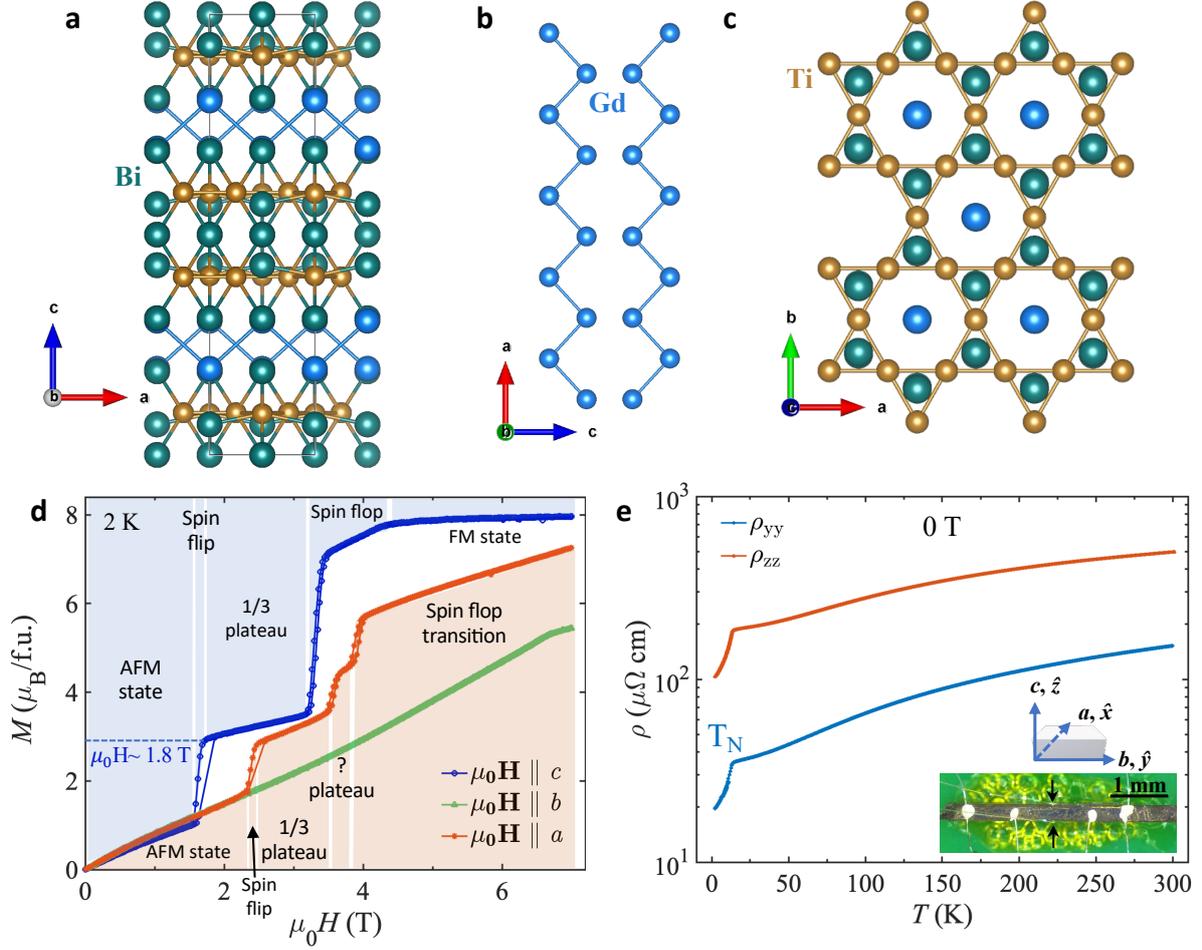

**Fig. 1.** (a) Crystal structure of GdTi$_3$Bi$_4$, (b) Zigzag chain of Gd atoms along *a*-axis, (c) Ti kagome lattice visualizing along *c*-axis, (d) Isothermal magnetization for $\mu_0 H \parallel a, b$ and $c$ at 2 K, (e) Temperature-dependent resistivity on log scale along $b$ ($\rho_{yy}$) and $c$ ($\rho_{zz}$) crystallographic axes at zero magnetic field. The inset shows an image of four-probe electrical contacts on the sample used for measuring resistivity and Hall voltage contacts shown using arrows.

High quality single crystals of GdTi$_3$Bi$_4$ were synthesized using Bi self-flux method.[29] Unlike other kagome prototypes *A*V$_3$Sb$_5$ and *Ln*M$_6$X$_6$, GdTi$_3$Bi$_4$ displays a more complex crystal structure (Fig. 1a).[20,22] It features a layered crystal structure with space group *Fmmm* and lattice parameters of $a$ = 5.858 Å, $b$ = 10.299 Å, and $c$ = 24.726 Å. The unit cell comprises a Ti-based kagome layer perpendicular to the *c*-axis and Gd-based zigzag chains along the *a*-axis. Interestingly, Ti-kagome layers show an additional anisotropic distortion, resulting in orthorhombic crystal symmetry rather than hexagonal. Fig. 1b emphasizes the presence of Gd-based zigzag chains aligned parallel to the *a*-axis, with nearest Gd−Gd spacing ~ 3.970 Å. Zigzag chains are fundamental motifs in low dimensional magnetism, as they typically host both comparable nearest-neighbour ($J_1$) and next nearest-neighbour ($J_2$) interactions, which introduces magnetic frustration.[43,44] Such systems exhibit magnetization plateaus at fractional values (e.g., 1/3) of the saturation magnetization and strongly influence the magnetoelectrical transport property of the material, as discussed in following sections.

After the successful growth of single crystals, the preliminary characterizations were carried out to confirm the composition, structure, and orientation (Fig. S1).[45] Temperature dependent field-cooled (FC) magnetic susceptibility ($\chi_c$) at 50 mT along the $\mu_0 H \parallel c$ exhibits a peak at approximately 14.73 K, indicating paramagnetic to AFM transition (Fig. S2a).[45] Fig. S2b-d shows isothermal magnetization measurements conducted along different axes and Fig. 1d shows isothermal magnetization along different directions at 2 K, with *a*- and *c*-axes being the easy axes. Magnetization for field along *c*-axis reaches saturation, $M_s$ ~ 7.9 $\mu_B$/f.u., which is in close alignment with the theoretical moment of the $Gd^{3+}$ ions. Notably, a distinct metamagnetic transition below 8 K is discernible at critical field of approximately $\mu_0 H_C \approx 1.8$ T (*c*-axis) and 2.5 T (*a*-axis). These transitions are at approximately 1/3 of $M_s$. Both directions show a spin-flip transition before 1/3$^{rd}$ plateau, followed by spin-flop. For $\mu_0 H \parallel a$, an additional plateau like behavior appears before going to FM state. We have termed it unknown plateau for this study. Overall magnetization varies similarly for both the $\mu_0 H \parallel a$ and $\mu_0 H \parallel c$, even though magnetization saturation is approaching trend till 7 T for *a*-axis. This similarity yet difference in magnetization hints at the direct link with zigzag chain of Gd along *a*-direction. In GdTi$_3$Bi$_4$, weak interlayer AFM coupling leads to the realisation of 1/3 plateau at comparatively low fields for $\mu_0 H \parallel c$. The presence of similar multiple metamagnetic transitions has been previously observed in HoAgGe as well and has been attributed to the existence of canted AFM states.[42,46] Such states exert a significant influence on transport properties, highlighting the importance of understanding the magnetic behavior. As sketched in the inset of fig.1e, the *a*-, *b*-, and *c*-axes of the material are equivalent to the $\hat{x}$, $\hat{y}$, and $\hat{z}$ in measurements, respectively.

To examine the complex relationship between magnetism and transport property, the measurement of transverse resistivity and Hall resistivity of GdTi$_3$Bi$_4$ crystals was carried out across 2-30 K under varying temperatures and $\mu_0 H \parallel \hat{z}$ and $\hat{x}$. Fig.1e shows zero-field resistivities ($\rho_{yy}, \rho_{zz}$) in logarithmic scale versus temperature. The decrease in resistivity with decreasing temperature is indicative of metallic behavior, while the hump near 14.7 K marks the paramagnetic to AFM transition. Below the magnetic transition, a sharp decrease in the resistivity indicates lower scattering, and this decline is particularly pronounced in the case of a transition to an AFM state. The resistivity along $\hat{z}$ has been found to be ~ 3 times higher in comparison to that of $\hat{y}$. This resistivity anisotropy indicates that GdTi$_3$Bi$_4$ is indeed a layered material, wherein layers are not highly isolated, and they are electrically connected.[47] These resistivities at different magnetic fields are shown in Fig. S3a-b, indicating the suppression of $T_N$ with field due to the predominance of Zeeman energy over antiferromagnetic exchange energy.[45]

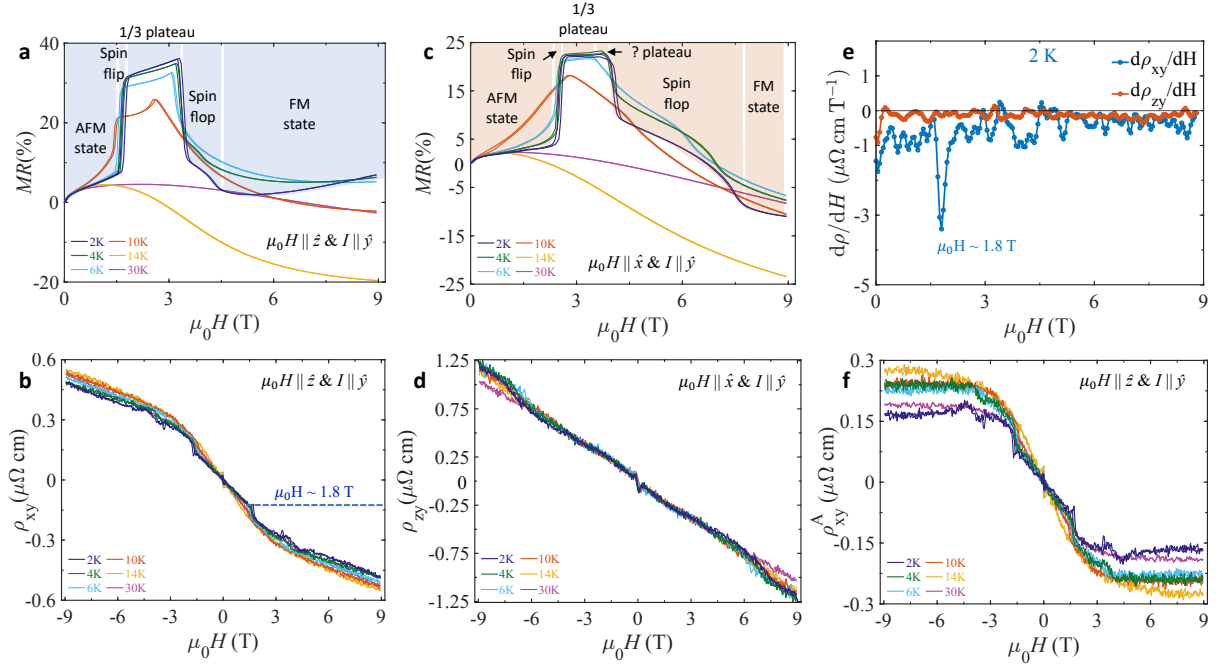

**Fig. 2.** Field-dependence of (a) resistivity and (b) Hall resistivity at various temperatures for $I \parallel \hat{y}$ and $\mu_0 H \parallel \hat{z}$, field-dependence of (c) resistivity and (d) Hall resistivity at various temperatures for $I \parallel \hat{y}$ and $\mu_0 H \parallel \hat{x}$, (e) comparison between first derivative of Hall resistivities of two different directions ($\rho_{zy}$ & $\rho_{xy}$) at 2 K, (f) extracted anomalous Hall resistivity ($\rho_{xy}^A$) from (b).

The application of $\mu_0 H \parallel \hat{z}$ and $\hat{x}$ enabled the discernment of AFM ground state–1/3 plateau spin-flip–FM spin-flop transitions from the field-dependent resistivity plots at different temperatures. The magnetoresistance (*MR*) data (*MR* (H) = [$\rho_{yy}$(H)- $\rho_{yy}$(0)]/ $\rho_{yy}$(0)), displays a complex behavior at low temperatures (Figs. 2a & 2c). Specifically, at 2 K for $\mu_0 H \parallel \hat{z}$, MR initially increases gradually before exhibiting sharp transitions at approximately 1.8 T, 3.3 T, and 4.4 T, which correspond to the metamagnetic transitions observed in the magnetization data. These abrupt shifts in resistivity can be attributed to the presence of canted spin states. These features are only evident below 15 K due the occurrence of spin-flip transitions. As the temperature increases, the peaks gradually shift to lower fields and eventually disappear above the $T_N$. The magnetization data corroborate these findings and offer a potential explanation that is rooted in metamagnetic phase transitions. A similar MR behavior was observed for $\mu_0 H \parallel \hat{x}$, as shown in Fig. 2c. It can be concluded that the magnetic property of the material is the primary factor influencing its electronic band structure and its topology, which can be accessed through the transport measurements.[5,40,48]

Fig. 2b presents the Hall resistivity at different temperatures (2 K to 30 K) as a function of the applied magnetic field, when $\mu_0 H \parallel \hat{z}$ and I $\parallel \hat{y}$. In magnetic systems, the Hall resistivity ($\rho_{xy}$) is typically composed of two distinct contributions: $\rho_{xy} = R_0\mu_0 H + R_S M$, where the first and second terms represent ordinary and anomalous Hall resistivities, respectively.[3] By isolating these contributions, we obtained the anomalous Hall resistivity ($\rho_{xy}^A$), as depicted in Fig. 2f. The $\rho_{xy}^A$ exhibits stepwise increments with $\mu_0 H$, followed by a saturation and subsequent increments, which mirror the behavior observed in the magnetization curve. After extrapolating the saturation to zero field at 2 K, the typical value of $\rho_{xy}^A$ is ~0.18 μΩ cm (Fig. 2f). In contrast, after switching the $\mu_0 H \parallel \hat{x}$ in the same piece of sample, the $\rho_{zy}$ shows completely linear behavior (Fig. 2d). This can be seen better in the first derivative of Hall resistivities with respect to the field for 2 K, as shown in Fig. 2e. The substantial alteration in slope is manifested as a dip when $\mu_0 H \parallel \hat{z}$ and this dip is absent when $\mu_0 H \parallel \hat{x}$. Interestingly, the field value of the dip corresponds to the 1/3 plateau, where anomalous behavior in the Hall resistivity exclusively emerges in the $\mu_0 H \parallel \hat{z}$ case. This feature has been thoroughly examined in subsequent discourse, accompanied by theoretical insights. In other study, the Hall signal has similar features except a pronounced peak at ~ 3T, which could be due to a microscopic origin.[30] Temperature dependent variation of both mobility ($\mu$) and carrier concentration ($n$) at different directions has been shown in Figs. S3c-d.[45] The comparable $\mu$ and $n$ along both axes indicate that underlying dominant scattering mechanisms are similar. The presence of a cusp-like feature in magnetoresistance and a kink in Hall resistivity for both field orientations near zero field is attributed to weak anti-localization effect, which is commonly observed in compounds with large SOC.[49,50]

The presence of anisotropic AHE wherein one direction and absence in another despite analogous magnetization is rarely observed in materials. To best of our knowledge, there is no extant experimental study that claim to observe this phenomenon. This alteration may be attributable to the intertwining of magnetism and the zigzag configuration of Gd atoms along the *a*- axis. Recent observations have revealed the presence of a directional-dependent spin-charge intertwined density wave and stripe magnetization in this material when an external magnetic field is applied along the *c*-axis direction.[30,32] A similar intertwining phenomenon has also been observed in another kagome material, $Mn_3Sn$, suggesting a potential link to the magnetization.[51] We have further elucidated the origin of observed AHE along *c*-axis as shown in Fig. 3.

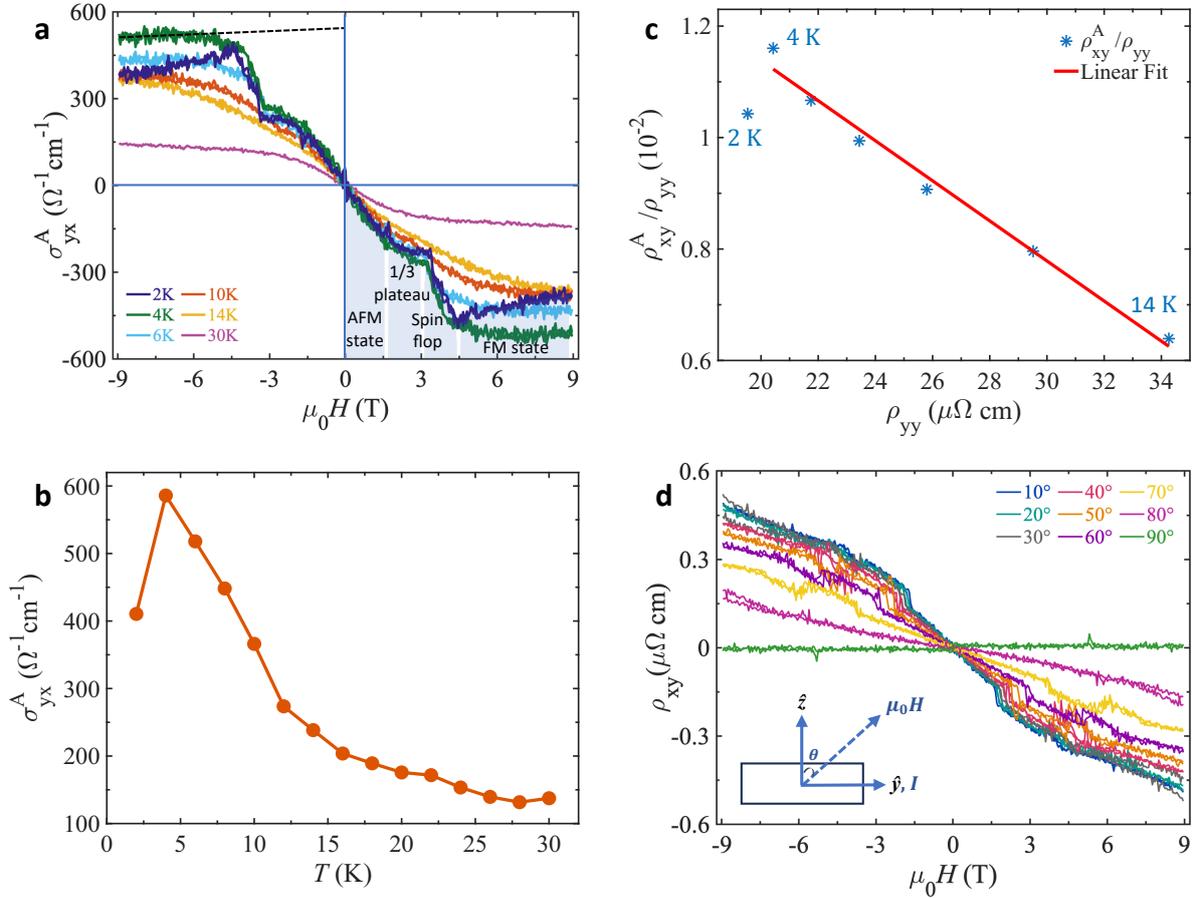

**Fig. 3.** (a) Anomalous Hall conductivity, $\sigma_{yx}^A$ of GdTi$_3$Bi$_4$ at several temperatures, (b) variation of $\sigma_{yx}^A$ with temperature, (c) linear fitting of the ratio of $\rho_{xy}^A/\rho_{yy}$ versus $\rho_{yy}$ in the temperature range 4 K - 14 K, (d) Hall resistivity in different angles at 2 K. The inset shows the rotating angle between $\mu_0 H$ and $I$.

The field dependent AHC ($\sigma_{yx}^A$) was estimated by employing the equation $\sigma_{yx}^A(H) = \frac{\rho_{xy}^A(H)}{(\rho_{xy}^A)^2(H)+\rho_{yy}^2(H)}$. As displayed in Fig. 3a, the calculated $\sigma_{yx}^A$ at 4 K yields a value of 585 $\Omega^{-1}$ cm$^{-1}$, extrapolated from high field. AHE is a quantum phenomenon that arises from multiple reasons, but three main reasons are: Berry-phase, skew scattering from disorder in the system, and side jump scattering due to spin orbit coupled impurities.[3] The Berry phase contribution is intrinsic, arising from band structure of the material. The skew scattering is theoretically proportional to the resistivity [$\rho_{xy}^A \propto \rho_{yy}$]. On the other hand, contribution of Berry phase and side jump scattering together is proportional to the square of resistivity [$\rho_{xy}^A \propto \rho_{yy}^2$]. Thus, AHE can be written as the sum of both dependences as $\rho_{xy}^A = a\rho_{yy} + b\rho_{yy}^2$. Linear fitting of $\rho_{xy}^A/\rho_{yy}$ versus $\rho_{yy}$ from 4 K to 14 K (Fig. 3c) confirms that the main contribution of AHE is from the BC and side jump mechanisms. The slope of the fit yields an intrinsic AHC of ~ 360 $\Omega^{-1}$ cm$^{-1}$, closely matching with the experimental value. Notably, the data at 2 K deviates from a linear fit. This deviation is indicative of a potential change in the underlying electronic structure, which may be attributed to a transition from 3Q charge order to 2Q charge order.[32] Furthermore, variation of AHC with temperature also clearly shows the rise from 2 K to 4K, followed by substantially decline at temperatures exceeding 15 K (Fig. 3b). This behavior shows that AHE is not only direction sensitive, its magnitude and scaling behavior is also sensitive to ordering parameters. The angle-dependent AHE demonstrates a decline in value with increasing angle, as shown in Fig. 3d. It is evident that the perpendicular component of the field with respect to the current is the sole contributing factor to the Hall resistivity. Consequently, as the angle between $\mu_0 H$ and $I$ decrease, the Hall resistivity decreases.

Moreover, full logarithmic plot of $\sigma_{yx}^A$ versus $\sigma_{yy}$ for a range of materials is shown in Fig. S5.[8,9,45,52-59] In general, the analysis reveals the presence of three distinct regimes in the classification of AHE contribution. In high conductive regime [$\sigma_{yy} > 10^6$ $\Omega^{-1}$cm$^{-1}$], the linear contribution to $\sigma_{yx}^A \sim \sigma_{yy}$ is dominated by skew scattering, while in bad metal regime [$\sigma_{yy} < 10^4$ $\Omega^{-1}$cm$^{-1}$], $\sigma_{yx}^A$ decreases with decreasing $\sigma_{yy}$ at a rate that is faster than linear. On the other hand, within intrinsic or scattering independent regime [$10^4$ $\Omega^{-1}$cm$^{-1}$ < $\sigma_{yy}$ < $10^6$ $\Omega^{-1}$cm$^{-1}$], $\sigma_{yx}^A$ is roughly independent of $\sigma_{yy}$.[3] Thus, it can be concluded from Fig. S5 that GdTi$_3$Bi$_4$ lies in the intrinsic regime, thereby signifying that the observed AHE is indicative of an intrinsic contribution such as BC, which depends on the band structure of the material.

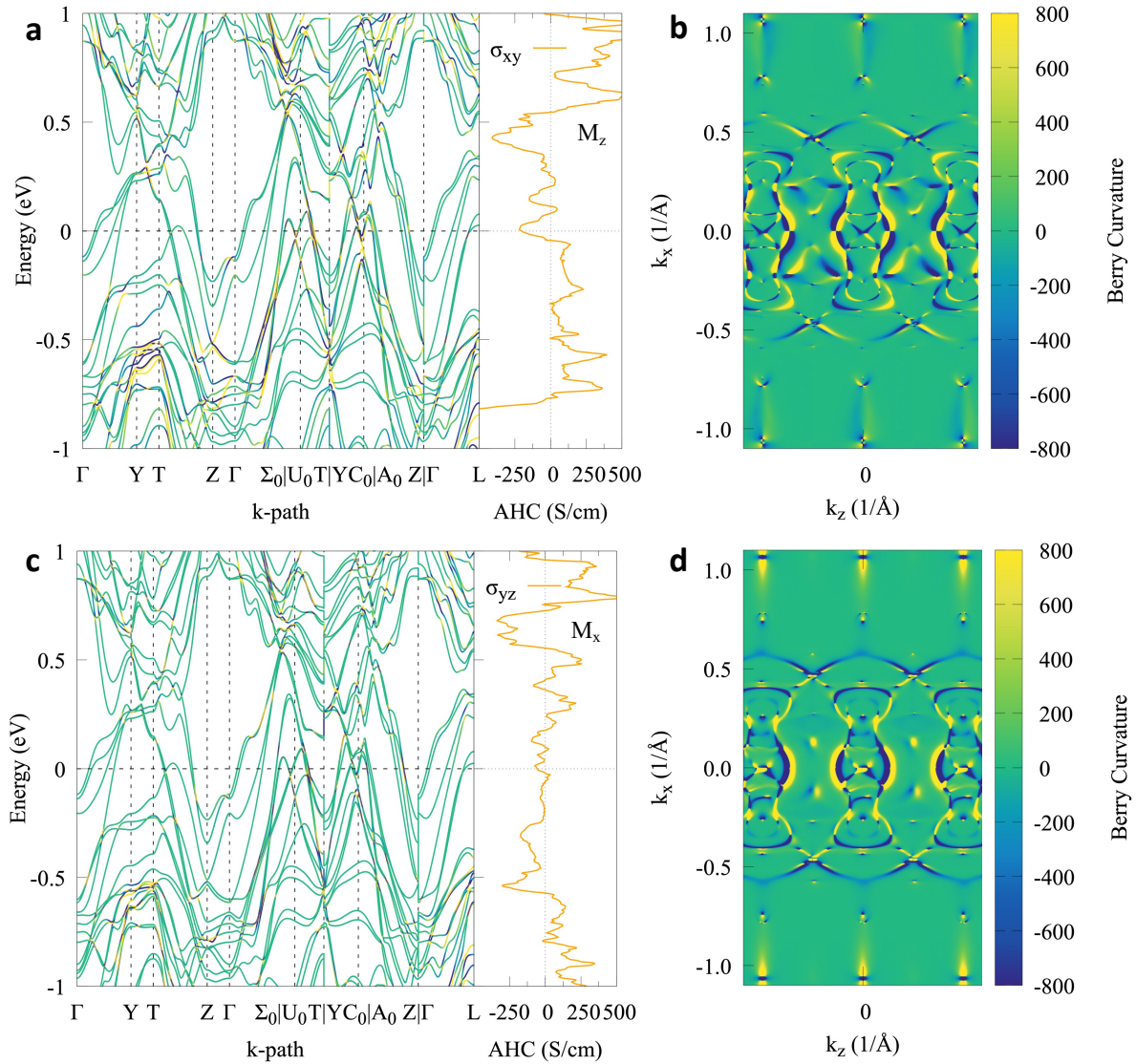

**Fig. 4.** (a) Berry-curvature–colored band structure for the saturated FM state with $\mu_0 H \parallel \hat{z}$ (left; bands colored by $W_z = \Omega_z$) and the intrinsic $\sigma^A_{xy}(E_F)$ from the Kubo–Berry formalism (right; $E_F$ indicated). (b) Constant-energy slice of $W_z(k)$ on the T-Y-U$_0$-C$_0$ plane at $E = E_F$. (c) Same as (a) but for $\mu_0 H \parallel \hat{x}$: left, bands colored by $W_x = \Omega_x$; right, intrinsic $\sigma^A_{yz}(E_F)$. (d) Same as (b) but for $\mu_0 H \parallel \hat{x}$: constant-energy slice of $W_x(k)$ on the T-Y-U$_0$-C$_0$ plane at $E = E_F$.

In the high-field saturated FM state, the TRS is broken, and the system carries a uniform non-zero net magnetization (M≠0) aligned with the external field, superimposed on an AFM background. The intrinsic AHC originates from BC and governed by the spin-quantization axis under SOC. To evaluate it in two saturated configurations, all local moments are constrained to be collinear with M ∥ a or M ∥ c (constrained-FM).[3,60,61] First-principles spin- polarized DFT calculations were performed using the Vienna ab initio simulation package with PBE-GGA and Dudarev DFT+U scheme (U$_{eff}$ = U−J = 6 eV for Gd-4f and all others = 0 eV).[62-65] Figs. 4a and 4c show the band structures for FM state along $\hat{z}$ and $\hat{x}$, respectively. The band colors encode the z- and x-components of the BC, Ω(k), on the color scale shown in Fig. 4b; yellow (blue) denotes large positive (negative) curvature. Right panels show energy dependent AHC obtained from Kubo- Berry formalism. At E = $E_f$, the calculated intrinsic AHC for $\mu_0 H \parallel \hat{z}$ is much larger than that for $\mu_0 H \parallel \hat{x}$ (by a factor of 3.6), and its sign and scale are comparable to the experimental AHC for $\mu_0 H \parallel \hat{z}$ demonstrating that the dominant contribution is intrinsic. BC slices (Fig. 4b and 4d) further highlight how the Hall response evolves when the magnetization is rotated. For $\mu_0 H \parallel \hat{z}$, Ω$_z$ on the T–Y–U$_0$–C$_0$ plane manifests both broader positive bands and deeper negative lobes, with negative region clearly predominating in terms of areal coverage (Fig. 4b). The Brillouin zone has been shown in Fig. S6.[45] Integrating the signed Berry curvature over this slice yields a net value of 169.75 (in units consistent with the 3D Kubo–Berry integration). In contrast, in Fig. 4d (Ω$_x$ on the same plane with $\mu_0 H \parallel \hat{x}$ ), the curvature generates narrow yellow- blue stripes of opposite sign and almost equal area; the net signed integral is only 16.62, indicating much stronger cancellation. Consequently, their contributions cancel in the Brillouin-zone integral and only a modest $\sigma_{yz}^A$ remains, where $\sigma_{yx}^A$ is set by an extended, nearly single-sign $\Omega_z$ pattern. By comparing orbital-projected band structures with Berry-colored bands, we find that for $\mu_0 H \parallel \hat{z}$ the BC hot spots near $E_F$ trace SOC-gapped anticrossings within the Ti $t_{2g}$ manifold, predominantly $\{d_{xz}, d_{yz}\}$ along Y–C$_0$ and U$_0$–T. In contrast, for $\mu_0 H \parallel \hat{x}$, the active SOC matrix elements couple $d_{xz}$ to $d_{z^2}$ and $d_{yz}$ to $d_{xy}$. However, the $d_{z^2}$ weight is shifted above $E_F$, resulting in the avoided crossings that fall away from $E_F$. This leads to a significantly weaker $\Omega_x$ and $\sigma_{yz}(E_F)$. Therefore, despite the negligible local moment of Ti, the exchange field from the Gd-4f sublattice breaks TRS for the Ti-derived Bloch states. In the presence of SOC, the active-matrix elements are found to be contingent upon the spin-quantization axis ($\mu_0 H \parallel \hat{x}$ vs $\mu_0 H \parallel \hat{z}$), elucidating the underlying reason for the substantial AHC for $\mu_0 H \parallel \hat{z}$ and its strong suppression for $\mu_0 H \parallel \hat{x}$.

**Conclusion:**

In summary, our study presents a systematic direction dependent investigation into the magnetic and electrical transport properties of GdTi$_3$Bi$_4$ single crystals. GdTi$_3$Bi$_4$ with a $T_N$ of approximately 15 K displays the metamagnetic transition under $\mu_0 H \parallel c$ and $a$, usually present in 1D zigzag chain systems due to weak interlayer AFM couplings. We observed that the large anomalous Hall conductivity of 410 $\Omega^{-1}$ cm$^{-1}$ at 2 K is observed when $\mu_0 H \parallel \hat{z}$ and it is absent when $\mu_0 H \parallel \hat{x}$ despite the presence of similar field-dependent magnetic behavior. This is an intrinsic AHC, which is further originated from the BC. The obtained results reveal a clear distinction in the roles of the different components. Gd-4f is responsible for breaking TRS establishing the magnetization direction, while the SOC in the Ti-derived bands converts orientation into BC and controls the intrinsic AHC. This study serves as a foundation for future research to clarify mechanistic breakdown of the AHE. Extending such investigations to other kagome materials, especially $Ln$Ti$_3$Bi$_4$ compounds has the potential to uncover additional exotic topological phenomena, wherein the interplay between topology, electron correlation, and magnetism assumes paramount important.

**Acknowledgements:**

A.K.S is thankful for discussions with S. Das and S. Chatterjee. This work was financially supported by the Deutsche Forschungsgemeinschaft (DFG) under SFB1143 (project no. 247310070), the Würzburg-Dresden Cluster of Excellence on Complexity, Topology and Dynamics in Quantum Matter—ctd.qmat (EXC 2147, project no. 390858490) and the QUAST-FOR5249-449872909.